\documentclass[preprint,11pt]{elsarticle}

\makeatletter
\def\ps@pprintTitle{%
 \let\@oddhead\@empty
 \let\@evenhead\@empty
 \def\@oddfoot{}%
 \let\@evenfoot\@oddfoot}
\makeatother

\usepackage[titletoc,title]{appendix}

\makeatletter
  \@addtoreset{chapter}{part}
  \@addtoreset{@ppsaveapp}{part}
\makeatother

\usepackage{array,graphicx}
\usepackage{pifont}
\usepackage{amsfonts}
\usepackage{amsmath,amsthm}
\usepackage{booktabs}
\usepackage[inline]{enumitem}
\usepackage{relsize}
\usepackage{algorithm}
\usepackage{algcompatible}
\usepackage{setspace}
\usepackage{xcolor}
\floatname{algorithm}{}
\let\Algorithm\algorithm
\renewcommand\algorithm[1][]{\Algorithm[#1]\setstretch{1.2}}
\usepackage{euscript}
\usepackage{enumerate}
\usepackage{enumitem}
\setlist{leftmargin=5.5mm}
\usepackage{subfigure}
\usepackage{verbatim} 
\usepackage{xspace} 
\usepackage{hyperref,soul}
\hypersetup{%
    pdfpagemode={UseOutlines},
    bookmarksopen,
    pdfstartview={FitH},
    colorlinks=false,
    linkcolor={blue},
    citecolor={red},
    urlcolor={blue}
  }
\usepackage{url}

\newtheorem{definition}{Definition}

\newtheorem{rulesH}{Rule}

\newcommand{\IQP}{$IQP$\xspace}

\definecolor{blue}{RGB}{59, 42, 247}

\begin{document}
\pagestyle{plain}

\begin{frontmatter}

  \title{On fairness and diversification in WTA\\and ATP tennis tournaments generation}
  \date{}

  \author[1,2]{Federico Della Croce}
  \author[1,3]{Gabriele Dragotto}
  \author[1]{Rosario Scatamacchia}

  \address[1]{\small Dipartimento di Ingegneria Gestionale e della Produzione, \\
   Politecnico di Torino, Italy. \\{\tt \{federico.dellacroce,rosario.scatamacchia\}@polito.it }}
  \address[2]{CNR, IEIIT, Torino, Italy}
  \address[3]{\small CERC Data Science for Real-time Decision-making, \\ \'Ecole Polytechnique de Montréal, Canada. \\{\tt gabriele.dragotto@polymtl.ca }}

  \begin{abstract}

Single-elimination (knockout) tournaments are the standard paradigm for both 
main tennis professional associations, WTA and ATP. Schedules are generated by 
allocating first seeded and then unseeded players with seeds 
prevented from encountering each other early in the competition. 
Besides, the distribution of pairings in the first round between 
unseeded players and seeds for a yearly season may be strongly 
unbalanced. This provides often a great disadvantage to some 
"unlucky" unseeded players in terms of money prizes. Also, a fair 
distribution of matches during a season would benefit from limiting 
in first rounds the presence of Head-to-Head (H2H) matches between 
players that met in the recent past. 

We propose a tournament generation approach in order to reduce in the first round unlucky pairings and also replays of H2H matches. The 
approach consists in a clustering optimization problem inducing a 
consequent draw within each cluster. A Non-Linear Mathematical 
Programming (NLMP) model is proposed for the clustering problem so as 
to reach a fair schedule. The solution reached by a commercial NLMP 
solver on the model is compared to the one reached by a faster hybrid 
algorithm based on multi-start local search. The approach is 
successfully tested on historical records from the recent Grand Slams 
tournaments. 

  \end{abstract}

\end{frontmatter}

\section{Introduction}
\label{sec:intro}

Algorithms and quantitative approaches are increasingly becoming a key aspect of the sports industry as discussed, e.g., in \cite{KENDALL20101}. The large number of stakeholders present in sports planning and scheduling creates favorable conditions for optimization-based approaches. In general, maximizing revenues and keeping sports games attractive for both media and fans are two of the most important aspects involved in scheduling sports competitions.
Also, athletes are mainly concerned with their career and correspondingly are interested in having a schedule that positively affects their performances and returns. We turn our attention, here, to tennis tournaments generation with a particular reference to professional tennis tournaments and the related associations, namely WTA for women and ATP for men.

The vast majority of professional tennis tournaments foresees a
single-elimination tournament where the loser of a match is directly eliminated from the tournament, while the winner moves on to the next round. The tournament ends when the two remaining players are opposed in the final match leading to a final winner.
Given the set of participants, a draw takes place among the players in order to generate the first-round brackets graph where players are split into two subsets, seeded players - the ones with highest rankings - and unseeded ones.  The first two seeded players usually have an a-priori allocated slot in the brackets graph, while the remaining seeds have a restricted set of slots in which they can be allocated. Hence, a constrained draw for seeds is made before the one for unseeded players. The seeding process ensures that the best players do not meet in the first rounds of the competition. Once the draw among seeds is established, a second draw takes places among the unseeded players in order to fill all the empty slots of the brackets graph in the first round.

We consider here the allocation mechanism for unseeded players, assuming that seeding has already been provided. 
We provide a fairness-based approach in order to ensure that the generated schedule fits 
additional requirements in terms of impartiality, fairness, and minimization of match replays between recent opponents.

We focus on WTA and ATP Grand Slams, the four most prestigious tennis tournaments in professional leagues. In such tournaments, most of the top-ranking players are competing. Correspondingly, these tournaments are the most appealing for both fans and sponsors and money prizes are the highest in the season.
As noted in \cite{forestetal,Dagaev2018}, the general interest in matches is directly related to the uncertainty of outcomes and competitive intensity between opponents. 
With respect to professional tennis tournaments, we may assume that the predictability of outcomes can also be influenced - to some extent - by the number of times two opponents played against each other. The more information is available about matches of two players (e.g., the so-called Head-to-Head  or H2H index), the more accurate predictions can be given about the outcome of a match between them. On the other side, apart from top players, such a match can turn out to be less appealing to the public, particularly if it occurs in the first rounds of the tournaments.

We propose an algorithmic approach with the aim of maximizing the diversification of pairings in the very first rounds and avoiding frequent match replays in those rounds. While rivalries among top players drive much of the interest in tennis and replays in the final tournaments rounds are what many supporters look for, match replays in the very first rounds are much less appealing, particularly between unseeded players.
Further, we focus on a phenomenon, more frequent than what one may expect, that is related to unseeded players 
that are repeatedly paired in the first round with seeded players. Hereafter, we will refer to those players as $u$-players, and a match between one of these players with a seed as a $u$-pairing.

We take also into account others parameters such as players nationality as potential elements of disparity in a schedule. 
Generally speaking, the cost of pairing can be extended to any other parameter of interest. For instance, when players get wild-cards, it might be of interest to penalize the pairing of this wild card in the first round with some given players.
The aim of the proposed approach is to create tournament schedules that minimize a generic pairing cost function. 
We propose an optimization approach where we cluster players into different groups in order to minimize the mutual pairing costs inside each group. A draw is then performed within each cluster. 
For the solution of the clustering phase, an {Integer Quadratic Programming} (\IQP) model is presented and applied to the above mentioned Grand Slam instances. For that phase, we also propose a two-step heuristic procedure capable of reaching good results within a very limited CPU time. The computational tests highlight how such an approach can turn into quantifiable benefits for both players and audience.

Single-elimination tournaments have been deeply studied in the fields of Statistics, Combinatorial Mathematics and Operations Research.
Most of the literature related to optimization in tennis actually focuses on round-robin tournaments (see, e.g., \cite{dcat}) without taking into account the fairness aspects addressed in this article.
An extensive relatively recent literature review on scheduling in sport is provided by \cite{KENDALL20101} and covers a wide range of optimization approaches and sports applications. 

In \cite{farmeral2007} a method for allocating umpire crews in professional tennis tournaments is proposed. In \cite{Dagaev2018}, 
the problem of finding optimal seedings in single-elimination tournaments in order to take into account the competitive intensity and quality of every match is analyzed. 
In  \cite*{horen1985comparing} a statistical work is proposed for single-elimination tournaments, pointing out how different brackets graphs lead to diverse patterns of winners and losers. According to that work, the tournament configuration can advantage or disadvantage contenders, therefore creating potential cases of iniquity. 
 In 
\cite{glickman2008}, a bayesian  optimal  design  approach  is proposed for single-elimination tournaments  that optimizes the probability that the best player wins in the current round.
The inpact of seeding procedures in terms of fairness is investigated in \cite{schwenk2000,karpov2016,karpov2018}. In \cite{williams2010fixing}, it is shown 
that - under certain assumptions - there is always a specific tournament structure which maximizes the odds of winning for any generic player. 
In \cite{hg16},  a methodology for finding globally optimal single-elimination tournament designs is proposed when partial information is known about the strengths of the players.
In \cite{acpr2017} the players winning probability in single-elimination tournaments is studied under several distinct assumptions. 
With respect to the literature, we propose a schedule generation approach which focuses on fairness in terms of repeated H2H matches and $u$-pairings, assuming that a seeding is given.

\section{Ensuring fairness and diversity}
\label{sec:fairness}

The success of a tennis player is strongly related to the rank in the leagues' leaderboards, drafted by the WTA and the ATP associations. A  professional career requires, among others, a strong economical effort. Professional tennis associations estimated that an average player traveling to 30 tournaments with a coach has to cover costs ranging from \$121.000 to \$197.000. On the other side, statistically, only the players ranked in the first 100 can cover such a cost. Therefore, according to \cite{reidetal2014}, being in the top 100 is not only a milestone in terms of recognition but a mandatory target for the development of a professional career. The unbalance between players actually making money and players struggling to break-even is a known problem in the professional tennis world (\cite{newman_2018}).
In the last years, several prize increase calls have been made from professional players (\cite{newman_2018}, \cite{gatto_2018}) and tournaments organizers are actually boosting economical rewards (\cite{bairner_2018}, \cite{maher_2017} and \cite{french_prize2018}).
Although prizes in the four Grand Slams have been increased by a 113\% in the last 10 years, most of the players outside the top 100 still struggle to cover the basic costs for their professional career (\cite{newman_2018}).

\begin{table}[htb]
  \centering
  \caption{Money prizes for winning $1^{st}$ and $2^{nd}$ rounds in the 2018 Grand Slams season (WTA and ATP). Adapted from \cite{french_prize2018}.\newline\newline}
  \label{unlucky:prizes}
  \begin{tabular}{@{}lcc@{}}
    \toprule
    \textbf{Tournament} & \textbf{$1^{st}$-round Prize} & \textbf{$2^{nd}$-round Prize}\\ \hline
    AUS     & \$48,000                  &  \$72,000\\
    ROL       & \$46,800                  & \$92,400\\
    WIM           & \$51,500                  & \$96,400\\
    US             & \$54,000                   &\$93,000\\ \hline
    Average             & \textbf{\$50,075}      & \textbf{\$88,450}  
  \end{tabular}
\end{table}

As shown in Table \ref{unlucky:prizes}, winning the first-round in a Grand Slam tournament can significantly impact the yearly income of an emerging tennis professional. If we take into account the average estimated yearly cost for a tennis professional (provided by \cite{reidetal2014}), a single first-round prize can cover from 23\% to 38\% of players costs. Reaching the second round of a Grand Slam tournament can nearly be the turning point into the career of a young player.

In general, unseeded players are expected to lose against seeded ones with high probability. Hence, it is crucial for them not to be paired to seeded players in the first round of Grand Slam tournaments.
However, historical data suggest that several unseeded players are paired - on first-rounds - with seeds in three or more Slam tournaments in a single season. 
We highlight how such situations can lead to significant damages in terms of career and prizes.
To this extent, we analyzed all Grand Slam tournaments for the seasons in years 2013-2018. 
Table \ref{table:unlucky_distribution} provides statistics on the number of times unseeded players are paired with seeds, on first-rounds, three or four times in a year. Also, the number of unseeded players (denoted TOT-U) participating to three or more Slams in the season is reported.
We note that, in years 2013-2018, TOT-U ranges both for ATP and WTA from 67 to 75.
\begin{table}[htb]
  \caption{Unlucky players for WTA and ATP seasons from 2013 to 2018.}
  \label{table:unlucky_distribution}
  \centering
  \begin{tabular}{@{}lccclccc@{}}
    \toprule
    ATP Season       & \textbf{3/4} & \multicolumn{1}{l}{\textbf{4/4}} & \textbf{TOT-U}        & WTA Season       & \multicolumn{1}{l}{\textbf{3/4}} & \multicolumn{1}{l}{\textbf{4/4}} & \textbf{TOT-U}        \\ \midrule
    ATP 2013         & 6            & 0                                & 75          & WTA 2013         & 7                                & 0                                & 75          \\
    ATP 2014         & 3            & 1                                & 71          & WTA 2014         & 8                                & 0                                & 72          \\
    ATP 2015         & 3            & 0                                & 71          & WTA 2015         & 8                                & 0                                & 74          \\
    ATP 2016         & 9            & 1                                & 69          & WTA 2016         & 3                                & 1                                & 71          \\
    ATP 2017         & 5            & 1                                & 68          & WTA 2017         & 8                                & 1                                & 75          \\
    ATP 2018         & 5            & 2                                & 68          & WTA 2018         & 9                                & 1                                & 67          \\
     \midrule
    \textbf{Average} & \textbf{5,2} & \textbf{0,8}                     & \textbf{70,2} & \textbf{Average} & \textbf{7,2}                     & \textbf{0,5}                     & \textbf{72,3} \\ \bottomrule
  \end{tabular}
\end{table}
In the considered time span, on average, 6 unseeded players were paired with a seed three times or more in ATP tournaments, while this entry increases to 7.7 for WTA tournaments (on average, approximately 8.6\% for ATP and 10.6\% for WTA). 
Although it might not be expected to have unseeded players paired with seeds in almost all the first-rounds of a single season, the evidence suggests that this phenomenon occurred quite often both in WTA and ATP Slams.

The real data of Table \ref{table:unlucky_distribution} show that the above mentioned players are far from being a theoretical speculation. 
Actually, given the money prizes reported in Table \ref{unlucky:prizes}, these players may suffer from a heavy economical damage and may correspondingly be affected by setbacks in their professional careers. Hereafter, we will refer to these players as {\em unlucky players } according to the following definition.
\begin{definition}
	An unlucky player is an unseeded player who is paired with a seeded player in the first round of three or four Grand Slam tournaments in a season. 
\end{definition}

\begin{figure}[!htb]
  \centering
  \includegraphics[width=\textwidth]{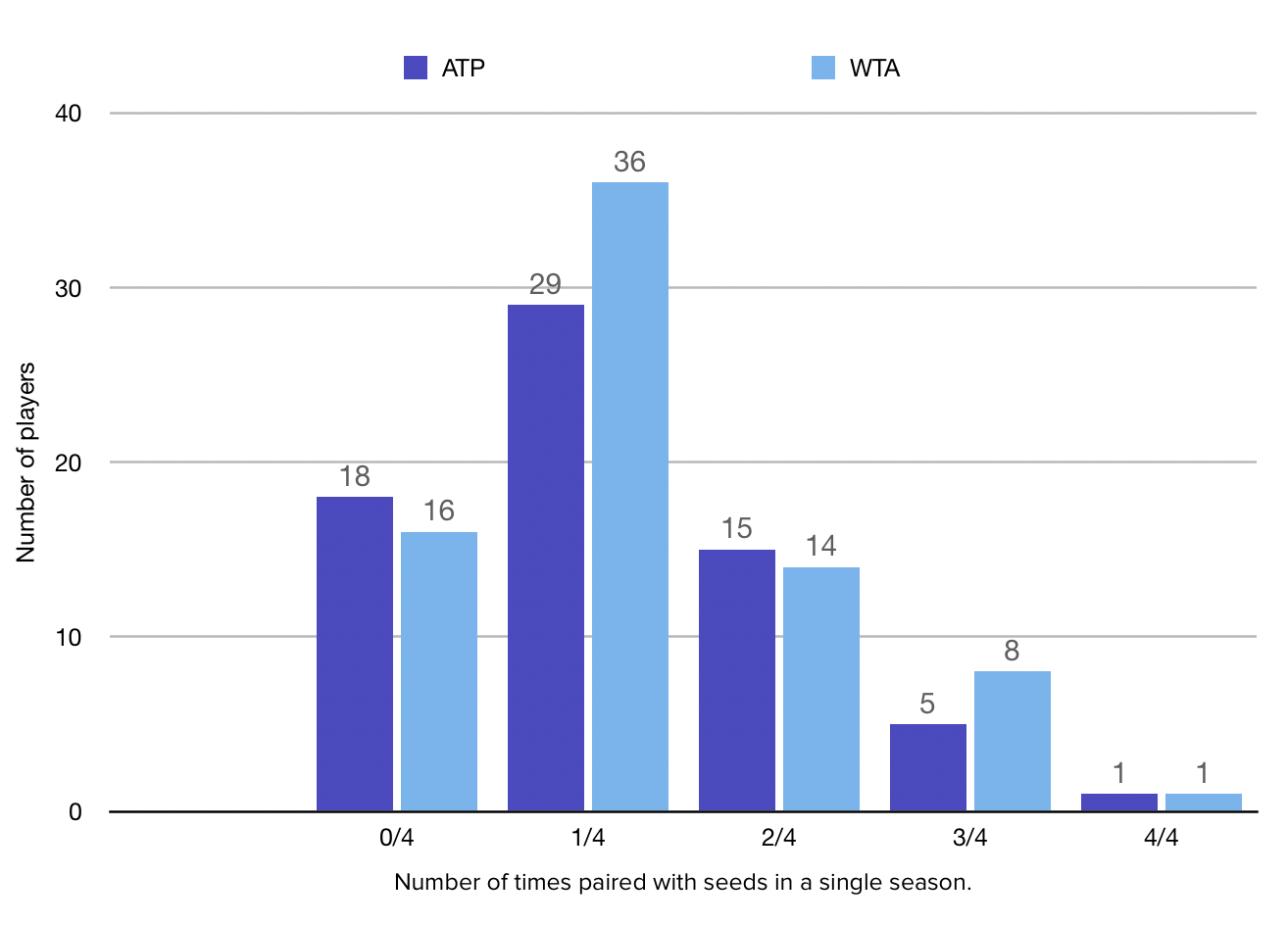}
  \caption{Distribution of pairings between unseeded players and seeds in the 2017 Grand Slam season for WTA and ATP.}
  \label{fig:unlucky_dist}
\end{figure}

By looking at the distribution of pairings between unseeded players and seeds for year 2017 in Figure \ref{fig:unlucky_dist}, we can easily spot the unbalance between the occurrences. In fact, many players have a limited number of pairings with seeds while some of them are unlucky.
While a strong correlation between unlucky pairings and prizes cannot be stated, the ranking positions of those players are generally negatively affected in both WTA and ATP. According to the argument provided in this section, a more balanced distribution of pairings between seeds and unseeded players can constitute a reasonable claim. Correspondingly, a primary aim is to generate schedules avoiding unlucky players.

\subsection{Diversity and pairing cost in the first round}
\label{sub:diversity}
With respect to the pairing of players in the first round of any given tournament as an outcome of the related draw,
having a diverse set of matches between players means avoiding frequent H2H matches that appeared in the past. This induces an increase in the number of different opponents a single player can have in the season. Nowadays, there are several cases in which players have been paired in the first round with the same opponent multiple times in a relatively small time span. We report some examples of frequent first-round pairings between players from the recent Grand Slams tournaments in Table \ref{table:FrequentPairings_2017}.
Extending this analysis to ATP and WTA 1000, 500 and 250 tournaments, there is a much larger evidence of this situation.
For instance, we checked the H2H activity in year 2018 of the ATP players that were ranked in positions 51-60 at the beginning of the year. All but two of them (Troicki and Benneteau who by the way had a reduced activity in that year) were paired more than once (in two cases three times) with the same opponent in the first round.

\begin{table}[htb]
  \centering
  \caption{Examples of frequent first-round pairings in recent Grand Slams for both WTA and ATP.}
  \label{table:FrequentPairings_2017}
\resizebox{\textwidth}{!}{%
  \begin{tabular}{@{}lllll@{}}
    \toprule
    \textbf{Tournament} & \textbf{Month/Year} & \textbf{Player A}  & \textbf{Player B}  & \textbf{League} \\ \midrule
    US            & Sept 2017            & Caroline Wozniacki & Mihaela Buzarnescu & WTA             \\
    AUS             & Jan 2018           & Caroline Wozniacki & Mihaela Buzarnescu & WTA             \\
    \midrule
    WIMB                & June  2017       & Elena Vesnina      & Anna Blinkova      & WTA             \\
    US             & Sept  2017         & Elena Vesnina      & Anna Blinkova      & WTA             \\ \midrule
    AUS            & Jan  2017          & Dudi Sela          & Marcel Granollers  & ATP             \\
    WIMB                & June  2017         & Dudi Sela          & Marcel Granollers  & ATP             \\ \midrule
    RG            & May  2018          & Nikoloz Basilashvili          & Gilles Simon  & ATP             \\
    WIMB                & June  2018         & Nikoloz Basilashvili          & Gilles Simon  & ATP             \\ \bottomrule
  \end{tabular}
  }
\end{table}

In terms of fairness, it makes sense to increase the probability of having first-round pairings between players that were never opposed. 
In terms of supporters attendance, other parameters such as the players nationality can be taken into account in the scheduling process
(it could be worthy, for instance, to avoid first-round matches between players of the same country). 
To this extent, we introduce the cost of pairing, so that a specific score can be attributed to each pair of players, and its value depends on the parameters of interest. This cost will be taken into account in the algorithmic approach described in the following section.

\section{Proposed approach}
\label{sub:twophase}
We consider a standard Grand Slam single-elimination tournament characterized by the following sets of players. The set $I :=\{i: 1\leq i\leq 128\}$ contains all the $n = 128$ players. 
The subset $M \in I$ has cardinality $m=32$ and contains seeded players, which are preventively assigned to standard predefined entries in the brackets graph. Then, a subset $U \in I$ {with cardinality $u=m=32$} contains the $u$-players in the previous 4 Grand Slam tournaments. More precisely, here the $u$-players are the set of unseeded players with the largest number of first-round matches with seeded players in such tournaments. The $u$-players cannot be paired with seeds, that is we avoid the presence of  $u$-pairings.
In order to maintain a draw procedure, as required in the generation of the first-round brackets graph for standard tennis tournaments, we propose the following approach.
We consider a clustering optimization problem, where the aim is to partition the players into $k=4$ different groups so that the pairing costs of the players assigned to the same cluster are minimized. 
The empirical evidence suggests that this number of clusters is suitable in order to achieve balanced outcomes while preserving a random draw inside sufficiently large clusters. 
The $u$-players are required to be uniformly split into each cluster ($u/k = 8$ players per cluster). Correspondingly, it will then be possible to have a draw within each cluster so that the pairing in the first round between $u$-players and seeds will be forbidden. Hence, the mutual costs between these players and the seeds are forced to $0$. Notice that, if clusters are generated as mentioned, a consequent draw can be executed in each cluster where, first, the pairings between the $m/k=8$ seeds and randomly selected players among the remaining $(128-m-u)/k=16$ players is generated and then a further draw (including this time the $u$-players) can be executed in order to generate the remaining pairings.
The rationale of this approach is to solve the clustering problem in order to facilitate fairness and diversification by minimizing the pairing costs between the players that will undergo the draw. 

\subsection{The clustering problem}
\label{sub:ClusteringPhase}

In order to minimize the players' pairing costs, a symmetric positive-defined $n \times n$ matrix $H$ is provided in input, where the generic element $h_{ \alpha \beta } \in H$ represents the pairing cost of two players $\alpha,\beta : \alpha,\beta \in I $. Notice that we pre-set $h_{\alpha\beta}=0\; \forall \; \alpha \in M, \beta \in U$, so that there is a zero cost between any seed $\alpha$ and $u$-player $\beta$ due to the fact that $u$-players will not be paired with seeds. As there are $k = 4$ clusters and each cluster will contain $n/k=32$ players with $m/k=8$ seeded players already predetermined, it follows that, in the clustering problem, we need to select for each cluster, $(128-m)/k=24$ unseeded players including $u/k=8$ $u$-players.

\subsubsection{Integer Quadratic Programming formulation}
\label{sub:IQP}
The clustering problem can be stated in terms of a quadratic 0/1 Mathematical Programming. We introduce a set of 0/1 variables $x_{ij}: i \in I, \; j \in J=\{1,...,4\}$ where $x_{ij} = 1$ if player $i$ is assigned to cluster $j$, $x_{ij} =0$ otherwise. Considering the pairing costs $h_{ \alpha \beta }$ introduced above, we obtain the following integer quadratic programming formulation.

\begin{align}
\min \qquad  & Z=\sum _{ j=1 }^{ k }{ (\sum _{ \alpha =1   }^{ n-1 } \sum _{ \beta =\alpha+1 }^{ n }{ h_{ \alpha \beta  }x_{ \alpha j } } x_{ \beta j } } ) & \label{eq:Objective} \\
s.t. \qquad  & \sum _{ j=1 }^{ 4 }{ x_{ ij } } =1 & \qquad \forall i\in I \label{eq:ConstraintMaxassignments}\\
& \sum _{ i=1 }^{ n }{ x_{ ij } } =n/k & \forall j\in J \label{eq:ConstraintMaxPlayerPerCluster} \\
& \sum _{ i \in U}{ x_{ ij } } =u/k & \forall j\in J \label{eq:ConstraintMaxUnluckyPerCluster}\\
& x_{ ij }  = 1 & \forall i \in M
\label{eq:SeededPlayersX}  \\
& x_{ ij }  \in \{0,1\} & \forall i \in I, j \in J \label{eq:DomainVariablesX}    
\end{align}
The objective function (\ref{eq:Objective}) minimizes the sum of pairing costs
of all pairs of players assigned to the same cluster. 
Constraints (\ref{eq:ConstraintMaxassignments})
require that every player must be assigned to one of the clusters,
while constraints (\ref{eq:ConstraintMaxPlayerPerCluster}) require that each cluster contains exactly $n/k$ players. Constraints (\ref{eq:ConstraintMaxUnluckyPerCluster}) guarantee that each cluster contains exactly $u/k$ $u$-players. Constraint (\ref{eq:SeededPlayersX}) fulfills the requirement on the pre-assigned seeded players. Finally, constraints (\ref{eq:DomainVariablesX}) indicate that the $x_{ij}$ variables are binary.

We remark that this problem is substantially equivalent (apart from the additional requirements on seeds and $u$-players and the minimization of the cost function) to the maximum diversity problem which is well known to be NP-Hard in the strong sense \cite{kuo1993analyzing}. 

\subsubsection{Heuristic solution of the clustering problem}
\label{sub:greedy}
Model \eqref{eq:Objective}-\eqref{eq:DomainVariablesX} can be solved by a nowadays commercial solver such as CPLEX. However, the quadratic nature of the problem may possibly affect the performance of a solver in providing good solutions in reasonable computational time. Also, in general, it is of interest to determine whether high quality heuristics may exist for a given combinatorial optimization problem.
In the light of these aspects, we also present a heuristic approach which provide instant feasible solutions to the clustering problem. The algorithm, denoted as $HEU$, provides solutions with an objective function very close to an optimal one (see Table \ref{table:ResultsSlams_2017} for numerical insights). We describe $HEU$ in the following, and provide the pseudo-code. 
We can represent the problem by means of a complete graph $G=(V,E)$, with set of vertices $V$ corresponding to the set of players, i.e. $V=I$, and set of edges $E$ where each edge $e_{ij}$ has a weight equal to entry $h_{ij}$ of matrix $H$. Correspondingly, each vertex $i$ has associated a weight $w_i$ equal to the weights of the edges emanating from it, namely $w_i = \sum_{j =1,...,n \cap j \not = i}h_{ij}$. Hence, nodes with a large weight correspond to players with a large amount of pairing costs. 
In the proposed approach, we first apply a greedy procedure (steps 2-8 of the pseudo code) that iteratively selects unseeded players one at a time in non-increasing order of weight $w_i$.
Then, the cluster for that player is determined. A cluster cannot be candidate for a player if $n/k$ players have already been assigned to that cluster. 
Likewise, as the number of $u$-players in each cluster is given, every time a $u$-player is considered, that player can be assigned to a cluster only if the number of $u$-players already assigned to that cluster is inferior to $u/k=8$. A selected player is assigned to the cluster $j_{min}$ that induces the least increase in the objective function value. If there are two clusters inducing the same increase, the one with the smallest index is selected. After a first solution is found, a simple local search procedure (steps 9-14) is launched as long as a time limit $T_l$ is not reached. 
Two different players $\alpha,\beta$ - respectively belonging to different clusters $j_{\alpha}$ and $j_{\beta}$ - are iteratively selected in a random way. The players can be both $u$-players or both unseeded. If swapping players $\alpha$ and $\beta$ by assigning them respectively to cluster $j_{\beta}$ and cluster $j_{\alpha}$ induces an improvement in the objective function (the corresponding variation is denoted as $\Delta S_{\alpha\beta}$), the swap is performed. This randomness implemented within a multi-start approach can also improve the unpredictability of the final schedule.

\begin{algorithm}[]
\label{Algorithm}
\begin{algorithmic}[1]
\small
\STATE \textbf{Input:} $H$ Matrix, $I,M,U$ sets {and time limit $T_l$}. 
\STATE {Order elements of $I$ by non-increasing $w_i$}

\FORALL{$i$ in $I$\textbackslash $M$} 
\STATE Determine the candidate cluster $j_{min}$ for player $i$ such that\\
\quad \quad $j_{min}$ contains less than $n/k$ players \\
\quad \quad {\bf if} $i \in U$ {\bf then} $j_{min}$ contains less than $u/k$ $u$-players 
\STATE {Assign $i$ to $j_{min}$}

\ENDFOR

\WHILE{{time limit $T_l$ is not reached}} 
\STATE Pick two random players $\alpha \neq \beta \in I\backslash M$, with $\alpha \in j_{\alpha}$ and $\beta \in j_{\beta}$ 
	\IF{$\Delta S_{\alpha\beta}<0$}
		\STATE{\textit{Swap}: assign $\alpha$ to $j_{\beta}$ and $\beta$ to $j_{\alpha}$}
	\ENDIF
\ENDWHILE
\end{algorithmic}
\caption{\textbf{Algorithm $HEU$ }}
\end{algorithm}

\section{Computational results}
\label{sec:results}
We considered the WTA and ATP database provided by \cite{atpdb} and sourced from the official websites of the two leagues. Computational tests consider the 2017 season for the four Grand Slam tournaments: Australian Open ($AUS$), Roland Garros ($ROL$), Wimbledon ($WIM$) and US Open ($US$).  In order to determine pairing costs $h_{ij}$ between pairs of players $(i,j)$, we took into account some features of interest discussed in the previous sections, for instance by penalizing matches between players of the same country.

\medskip

We considered all pairs of players $\alpha,\beta \in I$, 
such that $\alpha \in V$ and $\beta \in M$ and set $h_{\alpha \beta}=0$.
Similarly, we set $h_{\alpha \beta}=0$ if $\alpha,\beta \in I$ and $\alpha$ or $\beta$ is a qualified player (in Grand Slam tournaments the main draw foresees the presence of several - approx 5 - players selected from a qualifying round that is not yet finished at the time of the draw).
For the remaining pairs, given two players $\alpha,\beta$, the cost $h_{\alpha \beta}$ is initially set to $0$.
Then, the following set of rules is applied for increasing the value $h_{\alpha \beta}$ based on 
the results of the previous four Grand Slam tournaments.
Those rules constitute just a viable option for determining the $h_{\alpha \beta}$ coefficients, but different options could be clearly considered.

\begin{rulesH}\normalfont
    If two players $\alpha,\beta \in I$ played against each other in a $1^{st}$ round in the last 4 tournaments, then $h_{\alpha \beta}+=5$.
\end{rulesH}
\begin{rulesH}\normalfont
    If two players $\alpha, \beta \in I$ are from the same country, then $h_{\alpha \beta}+=5$.
\end{rulesH}
\begin{rulesH}\normalfont
    If two players $\alpha,\beta \in I$ played against each other in a $2^{nd}$ round in the last 4 tournaments, then $h_{\alpha \beta}+=2$.
\end{rulesH}
\begin{rulesH}\normalfont
    If two players $\alpha, \beta \in I$ played against each other in a $3^{rd}$ round in the last 4 tournaments, then $h_{\alpha \beta}+=1$.
\end{rulesH}
\begin{rulesH}\normalfont
    If two players $\alpha, \beta \in I$ played against each other either in quarter-final or semi-final rounds in the last 4 tournaments, then $h_{\alpha \beta}+=0.5$.
\end{rulesH}

The testing compares draws obtained after a clustering phase to the official draw.
The contribution emerging from tests is twofold: on one side, we show how our approach can lead to improvements - in terms of fairness and balance - compared to the official draw in the selected tournaments. On the other side, the computational tests provide indications on the effectiveness of the proposed heuristic in solving the clustering problem by comparing its performances with the ones of solver CPLEX 12.7 launched on model \eqref{eq:Objective}-\eqref{eq:DomainVariablesX}.
Computational tests were carried out on a \textit{3,5 GHz Intel Core i7} with $16GB$ of RAM. After preliminary testing, $T_{l}$ was set to $0.8$ seconds. This time limit showed up to be sufficient to reach a local minimum for steps 9-14 in Algorithm  $HEU$.
Table \ref{table:ResultsSlams_2017} provides the relevant  results. Here, we denote by $h$-pairing a pairing between two players $i,j$ inducing a cost $h_{ij} > 0$. \\

For each tournament, we compare (i) the actual draw ($REAL$)  sourced from the official tournament bracket graph, 
(ii) a simulated draw which is repeated $100$ times ($REAL100$) and is based on the current rules for the tournaments draw generation,
(iii) the draw computed by first launching CPLEX 12.7 on model \eqref{eq:Objective}-\eqref{eq:DomainVariablesX} and then simulating the
draw in each cluster ($CPLEX$) and (iv) the draw computed after $100$ different runs of the heuristic algorithm ($HEU$). In each run of the heuristic procedure, given the clustering solution and corresponding fixed placement of the seeded players, a one-shot random placement of the unseeded players in the tournament brackets graph is executed. In this placement, first the unseed players ($u$-players excluded) are paired to seeds and then the other pairings are randomly determined. With respect to $CPLEX$, we remark that CPLEX always reaches the optimal solution value of the clustering problem and, given the clustering solution, $100$ simulations like the ones used for $HEU$ are applied. The entries in Table \ref{table:ResultsSlams_2017} are as follows. In column $1$ are depicted the selected competitions. In column $2$, we report the CPU time required to generate the clustering solution (for $CPLEX$ and $HEU$). For $HEU$,
the CPU time is the average value obtained from the 100 runs. Column $3$ provides the value of the objective function ($O.F.$) of model \eqref{eq:Objective}-\eqref{eq:DomainVariablesX} related to the clustering problem. 
For entries $REAL$ and $REAL100$, the clustering is induced by assigning the first $32$ players of the tournament brackets graph to cluster $1$, the second $32$ players of the tournament brackets graph to cluster $2$ and so on. Column $4$ provides average, minimum and maximum number (in the relevant cases) of $u$-pairings. Finally, column $5$ provides average, minimum and maximum value for the sum of $u$-pairings and $h$-pairings. It is noteworthy to point out that algorithm $HEU$ has performances - in terms of $O.F.$ - comparable to the ones of CPLEX, while the CPU times required by the heuristic are dramatically smaller. Also, we remark that the proposed approach provides strongly reduced pairing costs together with no $u$-pairings, such that a much more balanced tournament is obtained. Indeed, the results show that both for $CPLEX$ and $HEU$ the  sum of $u$-pairings and $h$-pairings is typically  around $1$ or $2$ units on the average indicating that, by means of this clustering and corresponding draw, it is possible to get a first round reasonably fair and diversified.

\begin{table}[!ht]
\caption{Computational results for 2017 season of Grand Slams}
  \label{table:ResultsSlams_2017}
\resizebox{\textwidth}{!}{
\begin{tabular}{@{}|r|r|r|r|r|r}
\toprule
\multicolumn{1}{|l|}{\textbf{} } & \textbf{Time}  & \textbf{O.F. Value}    &\textbf{\small {$u$-pairings}}   & \textbf{\small {$(u+h)$-pairings} }   \\ 
\multicolumn{1}{|l|}{\textbf{} } &                      & \textbf{avg (min-max)}    &\textbf{avg (min-max)}   & \textbf{avg (min-max)}   \\ 
\midrule
\textbf{WTA-AUS 2017} &  &  &  & \\
\emph{REAL} & ---& 565.00 & 14 & 22 \\
\emph{REAL100} & ---& 512.21 (413.5 - 681.0) & 9.84 (5.0 - 17.0) & 13.81 (8.0 - 23.0)\\
\emph{CPLEX} & 187.37 & 251.50 & 0.00 (---) & 1.77 (0.0 - 6.0)\\
\emph{HEU} & 0.75 & 260.67  (258.5 - 263.5) & 0.00 (---) & 1.71 (0.0 - 7.0)\\
\textbf{WTA-ROL 2017} &  &  &  & \\
\emph{REAL} & ---& 522.00 & 15 & 16 \\
\emph{REAL100} & ---& 439.37 (318.0 - 607.0) & 10.02 (3.0 - 19.0) & 14.30 (8.0 - 22.0)\\
\emph{CPLEX} & 45.48 & 229.00 & 0.00 (---) & 2.33 (1.0 - 5.0)\\
\emph{HEU} & 0.74 & 240.91  (240.0 - 243.0) & 0.00 (---) & 2.58 (1.0 - 6.0)\\
\textbf{WTA-WIM 2017} &  &  &  & \\
\emph{REAL} & ---& 474.00 &  15 & 19\\
\emph{REAL100} & ---& 402.75 (299.0 - 558.0) & 10.08 (6.0 - 15.0) & 13.27 (7.0 - 20.0)\\
\emph{CPLEX} & 3.43 & 176.00 & 0.00 (---) & 1.26 (0.0 - 5.0)\\
\emph{HEU} & 0.75 & 196.75  (190.0 - 201.0) & 0.00 (---) & 1.56 (0.0 - 5.0)\\
\textbf{WTA-US 2017} &  &  &  & \\
\emph{REAL} & ---& 693.00 & 15  & 20 \\
\emph{REAL100} & ---& 585.69 (463.5 - 768.0) & 10.03 (4.0 - 16.0) & 14.92 (5.0 - 25.0)\\
\emph{CPLEX} & 601.28 & 378.00 & 0.00 (---) & 2.79 (0.0 - 6.0)\\
\emph{HEU} & 0.74 & 387.41  (386.5 - 388.5) & 0.00 (---) & 2.71 (0.0 - 6.0)\\
\textbf{ATP-AUS 2017} &  &  &  & \\
\emph{REAL} & ---& 377.50 &  8 & 10 \\
\emph{REAL100} & ---& 353.33 (226.5 - 514.5) & 10.31 (5.0 - 17.0) & 12.87 (6.0 - 24.0)\\
\emph{CPLEX} & 2.93 & 151.50 & 0.00 (---) & 0.68 (0.0 - 4.0)\\
\emph{HEU} & 0.75 & 164.91 (161.5 - 166.5) & 0.00 (---) & 1.25 (0.0 - 4.0)\\
\textbf{ATP-ROL 2017} &  &  &  & \\
\emph{REAL} & ---& 386.50 &  16 & 16 \\
\emph{REAL100} & ---& 403.11 (262.5 - 568.5) & 9.49 (1.0 - 18.0) & 12.62 (6.0 - 22.0)\\
\emph{CPLEX} & 2.99 & 208.50 & 0.00 (---) & 0.99 (0.0 - 4.0)\\
\emph{HEU} & 0.75 & 219.33  (217.5 - 219.5) & 0.00 (---) & 1.43 (0.0 - 5.0)\\
\textbf{ATP-WIM 2017} &  &  &  & \\
\emph{REAL} & ---& 302.50 &  16 & 17\\
\emph{REAL100} & ---& 311.68 (223.5 - 420.5) & 9.92 (5.0 - 16.0) & 12.16 (6.0 - 20.0)\\
\emph{CPLEX} & 1.90 & 128.50 & 0.00 (---) & 0.77 (0.0 - 3.0)\\
\emph{HEU} & 0.75 & 137.33  (136.5 - 137.5) & 0.00 (---) & 0.96 (0.0 - 4.0)\\
\textbf{ATP-US 2017} &  &  &  & \\
\emph{REAL} & ---& 466.00 &  12 & 14 \\
\emph{REAL100} & ---& 390.79 (272.0 - 543.0) & 10.03 (3.0 - 16.0) & 13.07 (6.0 - 19.0)\\
\emph{CPLEX} & 3.92 & 190.00 & 0.00 (---) & 0.78 (0.0 - 3.0)\\
\emph{HEU} & 0.77 & 194.33  (192.0 - 197.5) & 0.00 (---) & 0.78 (0.0 - 3.0)\\
\bottomrule
\end{tabular}
}
\end{table}

In Table \ref{table:HeuristicsSlams_2017}, we report some further statistics for algorithm $HEU$. The results are averaged over the $100$ runs considered. The first column reports the percentage improvement in the objective function achieved by the local search. The second and third column are the attempted swaps and successful ones, respectively. The fourth column reports the average number of $h$-pairings in the first round, while the last column sums up the values of such pairings.
 
\begin{table}[!ht]
\caption{Additional statistics on the Heuristic for 2017 season of Grand Slams}
  \label{table:HeuristicsSlams_2017}
\resizebox{\textwidth}{!}{
   \begin{tabular}{@{}r|r|r|r|r|r}
\toprule
\textbf{}   & \textbf{Avg.$\Delta\%$}            & \textbf{Swaps} &  & \textbf{{\textbf{{$h$-pairings}}}} & {\textbf{Costs of {$h$-pairings} }}   \\ 
 &  & Attempted  &  Successful &  & \\ \midrule
WTA-AUS 2017 &                              &                     &                               \\
\textbf{HEU}           & -8.95  & 37254.5 & 11.0 &  1.71                         & 7.36            \\
WTA-ROL 2017   &                              &                     &                              \\
\textbf{HEU}           & -1.67  &  45050.5 & 4.50 & 2.58                         & 12.19            \\
WTA-WIM 2017       &                              &                     &                         \\
\textbf{HEU}          & -2.62   &  55100.5 & 4.0 & 1.56                         & 6.74              \\
WTA-US 2017         &                              &                     &                           \\
\textbf{HEU}          & -4.17   &  54476.5 & 10.0 & 2.71                         & 12.30            \\ 
ATP-AUS 2017 &                              &                     &                                   \\
\textbf{HEU}          & -0.50   & 16137.5 & 2.0  & 1.25                         & 4.51              \\
ATP-ROL 2017   &                              &                     &                                  \\
\textbf{HEU}          & -2.37   &  58020.0 & 3.0    & 1.43                         & 6.19             \\
ATP-WIM 2017       &                              &                     &                             \\
\textbf{HEU}           & -1.60  & 61874.0 & 2.5  &  0.96                         & 3.84               \\
ATP-US 2017         &                              &                     &                               \\
\textbf{HEU}          & -2.92   & 57775.5 & 6.0  & 0.78                         & 3.08                 \\ \bottomrule
\end{tabular}
}
\end{table}

From Table  \ref{table:HeuristicsSlams_2017} we evince that the number of successful swaps is limited compared to the attempted ones even though the successful swaps are quite efficient. Indeed,
the local search step in the heuristic is quite profitable as it decreases the objective function value by roughly $3.1\%$ on the average with respect to the greedy solution.
Also, the cost of the $h$-pairings after the simulation remains very limited.

\section{Conclusions}
\label{sec:conclusions}
The aim of this work has been to integrate concepts of fairness and balance - typically studied in other disciplines - with a combinatorial approach typical of OR. This cross-fertilization between disciplines led to an approach capable of implementing a concept of fairness in sports scheduling.
The initial driver of this work concerns the presence of unbalance in professional tennis competitions draws generation. As the practical evidence shows,
 the need of better approaches is quite evident and Operations Research can positively contribute to their development.
Indeed, the data reported from the literature and media suggest that purely random draws and prizes increases are not enough to cope with the growing financial disparity in tennis. 
With this paper, we aim to provide a practical way for measuring and improving diversity and fairness in tennis tournaments. A simple, instant and manual step in this direction would be to modify all Slam tournaments draws as follows: `` Select first the players to be paired to seeds without taking into account those players that in the previous Slam were paired to a seed in the first round. Then, conclude the draw as usual''. In this way, any player will never be paired in the first round to seeds for two consecutive Slams.

\bigskip

\subsection*{Code}
The full code and data is available online on gitHub at:\\
\url{https://github.com/ALCO-PoliTO/TournamentAllocationProblem}

\newpage
\subsection*{Acknowledgments}
The very pertinent remarks and suggestions of two anonymous reviewers are gratefully acknowledged. This work has been partially supported by "Ministero dell'Istruzione, dell'Universit\`{a} e della Ricerca"
Award "TESUN-83486178370409 finanziamento dipartimenti di eccellenza CAP. 1694 TIT. 
232 ART. 6".

\bigskip

\bibliographystyle{elsarticle-harv}
\bibliography{Biblio3}
\label{sec:bib}

\end{document}